\documentclass[10pt]{article}
\usepackage[dvips]{graphicx}

\usepackage{epsfig}
\usepackage{graphicx}
\usepackage{indentfirst}
\usepackage{amsmath}
\usepackage{amsfonts}
\usepackage{amssymb}

\begin{document}

\begin{center}
{\Large\bf  Reconstruction of $f(T)$ gravity according to  holographic dark energy \\}
\medskip
  M. Hamani Daouda $^{(a)}$\footnote{E-mail address:
daoudah8@yahoo.fr}\ ,
  Manuel E. Rodrigues $^{(a)}$\footnote{E-mail
address: esialg@gmail.com}\ and
    M. J. S. Houndjo $^{(b)}$\footnote{E-mail address:
sthoundjo@yahoo.fr} \vskip 4mm

(a) \ Universidade Federal do Esp\'{\i}rito Santo \\
Centro de Ci\^{e}ncias
Exatas - Departamento de F\'{\i}sica\\
Av. Fernando Ferrari s/n - Campus de Goiabeiras\\ CEP29075-910 -
Vit\'{o}ria/ES, Brazil \\
(b) ICRA - Centro Brasileiro de Pesquisas F\'{i}sicas - CBPF\\ Rua Dr. Xavier Sigaud, 150, Urca, 22290-180, Rio de Janeiro, Brazil\\
\vskip 2mm

\begin{abstract}
We develop the reconstruction of $f(T)$ gravity model according to the holographic dark energy. $T$ is the torsion scalar and its initial value from the Teleparallel gravity is imposed for fitting the initial value of the function $f(T)$. The evolutionary nature of the holographic dark energy is essentially based on two important parameters, $\Omega_V$ and $\omega_V$, respectively, the dimensionless dark energy and the  parameter of the equation of state, related to the holographic dark energy. The result shows a polynomial function for $f(T)$, and we also observe that, when $\Omega_V\rightarrow 1$ at the future time, $\omega_V$ may cross $-1$ for some values of the input parameter $b$. Another interesting aspect of the obtained model is that it provides the unification scenario of dark matter with dark energy.
\end{abstract}
\end{center}
Pacs numbers: 98.80.-k, 95.36.+x, 04.50.Kd

\section{Introduction}
The recent observational data implies that the current expansion of the universe is accelerating \cite{bamba1,bamba2}. Recent observations from Ia Supernovae, the Large Scale Structure and Cosmic Microwave Background anisotropies confirm this late time acceleration \cite{bamba1,bamba2,setareh2}. Many theories are developed in order to explain how this acceleration occurs. In the framework of the General Relativity, it is quite accepted that the responsible for this acceleration is the dark energy, with negative pressure, and little is known about its origin and nature until now. Among other, the simplest candidate for dark energy is the cosmological constant or the vacuum energy. Despite its agreement with the observational data, this model is facing serious problem of cosmological constant \cite{chinois5}. This problem arises mainly because the vacuum energy is considered in the context of quantum field theory in Minkowski background. However, as it is very well known, at cosmological scales, quantum effects of gravity may be taken into account,  and the above description of the vacuum energy would break. Thus, it is estimated that the correct theoretical value of the vacuum energy will be provided by a complete theory of quantum gravity. Although, we are lacking a complete theory of quantum gravity, but we can make some attempts to probe the nature of dark energy according to some principles of quantum gravity. The holographic principle \cite{chinois6} is an important aspect that may explain some problems of the cosmological constant and also those of dark energy. Due to the gravity, and according to this principle,  the number of degree of freedom of a local quantum field theory is related to the area of its boundary, rather than the volume of the system as expected when gravity is absent. Currently, an interesting attempt to achieve a good comprehension of the nature of the dark energy in the context of quantum gravity is called ``holographic dark energy" proposal \cite{setareh6,setareh7,setareh8,setareh9}. It is well known that the holographic principle is one of the important results of recent investigations that arise exploring the quantum gravity or the string theory \cite{chinois6}. In this sense, Cohen et al. \cite{setareh6} proposed an entanglement relationship between the  infra-red ($IR$) and ultraviolet ($UV$) cutt-offs due to the limitation set by the formation of a black hole, which establishes an upper limit for the vacuum energy, $L^3\rho_V\leq LM^2_p$, where $\rho_V$ is the holographic dark energy density related to the $UV$ 
cut-off, $L$ the $IR$ cut-off, and  $M_p$ the reduced Planck mass. Holographic dark energy looks reasonable, since it may provide simultaneously natural solutions to both dark energy and cosmological problems as shown  in \cite{setareh9}. \par
The introduction of a new component of dark energy to the whole energy budget is one way to investigate this mystery, but there are also other promising ways without the need of new forms of energy. It is also considered  the modification of the Einstein-Hilbert action  for large scales with higher order curvature invariant terms such as $R^2$, $R_{\mu\nu}R^{\mu\nu}$,  $R_{\mu\nu\lambda\sigma}R^{\mu\nu\lambda\sigma}$, or $R\Box^{k}R$ and also non-minimally coupled scalar field with term $\phi^2 R$. Among the most general modified  gravity, the well known and usual used is $f(R)$ gravity. Many works have been done in the order to better explain some characteristics of holographic dark energy both in framework of General Relativity and $f(R)$ theory  \cite{setareh19}$-$\cite{oliver35}.\par 
However, General Relativity and its modifications are not the only theories with which holographic dark energy can be explored. Recently, special attention has been attached to a theory known as $f(T)$ gravity, where $T$ is the torsion scalar. $f(T)$ gravity is a modification of the teleparallel gravity, for which just the torsion scalar is the gravitational part of the action. Moller and other authors undertook and developed the proposal for an analogy between the teleparallel gravity and the General Relativity \cite{daouda1ref2}. Models based on $f(T)$ gravity were investigated, in one hand, as an alternative to inflationary models \cite{setareft16,setareft17}, and on the other hand, as an alternative to dark energy models \cite{setareft18}. Recently, still in the framework of $f(T)$ gravity, new spherically symmetric solutions of black holes and wormholes are obtained in \cite{daouda1,daouda2}. Power-law solutions of the scale factor are investigated when the universe enters a phantom phase  and it is shown that such power-law solutions may exist as cosmological model in the framework of $f(T)$ gravity. \par
Recently, still in the framework of  $f(T)$ gravity, the cosmological evolutions of the equation of state for dark energy in the exponential and logarithmic as well as their combination theories have been extensively considered by K. Bamba et al \cite{arbitrebamba1}. They shown that the crossing of the phantom divide line can be realized in the combined f(T) theory even though it cannot be in the exponential or logarithmic $f(T)$ theory. In particular, they found that the crossing is from the parameter of the equation of state related to dark energy $\omega_{DE}>-1$ to $\omega_{DE}<-1$, in the opposite manner from $f(R)$ gravity models. They also demonstrated  that this feature is favored by the recent observational data. Moreover, K. Bamba et al demonstrated that in the exponential $f(T)$ model studied in \cite{arbitrebamba2}, it is impossible to have the crossing of the phantom divide line, $\omega_{DE}=-1$.\par 
In this paper, we perform the reconstruction of $f(T)$ gravity without resorting to any additional dark energy, that is, considering that the holographic dark energy is effectively described by the modification of the gravity with respect to the teleparallel gravity. Then, we use the holographic model of dark energy in spatially flat FRW universe, and obtain equation of state for holographic dark energy density in framework of $f(T)$ gravity. The function $f$ is decomposed as the sum of the teleparallel gravity plus a term characterized by a function $g(T)$. Differential equation of the function $g$ is established and solved directly leading to the $f(T)$ model coming from holographic dark energy, at least in the case where the universe is essentially dominated by holographic dark energy. Moreover, we observe that for some values of the input parameter $b$, $\omega_V$ may cross $-1$. Another interesting aspect of the obtained model is that it is suitable for explaining the unification of the dark matter with dark energy.\par

The paper is organized as follows. In Section $2$ a generality of $f(T)$ gravity is presented, the flat  metric of FRW is assumed and the holographic dark energy is introduced. In Section $3$ the conclusion and perspectives are presented.

\section{$f(T)$ gravity with the account of holographic dark energy}
We propose in this section to generalize the teleparallel Lagrangian $T$ to a function $f(T)=T+g(T)$, which is similar to the generalization of the Ricci scalar in Einstein-Hilbert action to the modified $f(R)$ gravity. We can then write the  action of $f(T)$ gravity, coupled with matter
$L_m$ by \cite{setareft18,setareft19,setareft20,setareft,
setareparticle}
\begin{equation}\label{manuel1}
S=\frac{1}{16\pi G}\int d^4x e \left[T+g(T)+L_m\right]_,
\end{equation}
where $e=det(e^i_{\mu})=\sqrt{-g}$. In what follows, we will assume the units $8\pi G=1$. Here, The teleparallel Lagrangian
$T$, known as the torsion scalar,  is defined as follows
\begin{equation}\label{manuel2}
T=S^{\:\:\:\mu \nu}_{\rho} T_{\:\:\:\mu \nu}^{\rho},
\end{equation}
where
\begin{eqnarray}
T_{\:\:\:\mu \nu}^{\rho}=e_i^{\rho}(\partial_{\mu}
e^i_{\nu}-\partial_{\nu} e^i_{\mu}),\label{manuel3}\\
S^{\:\:\:\mu \nu}_{\rho}=\frac{1}{2}(K^{\mu
\nu}_{\:\:\:\:\:\rho}+\delta^{\mu}_{\rho} T^{\theta
\nu}_{\:\:\:\theta}-\delta^{\nu}_{\rho} T^{\theta
\mu}_{\:\:\:\theta}),\label{manuel4}
\end{eqnarray}
and $K^{\mu \nu}_{\:\:\:\:\:\rho}$ is the contorsion tensor
\begin{eqnarray}
K^{\mu \nu}_{\:\:\:\:\:\rho}=-\frac{1}{2}(T^{\mu
\nu}_{\:\:\:\:\:\rho}-T^{\nu \mu}_{\:\:\:\:\:\rho}-T^{\:\:\:\mu
\nu}_{\rho}).\label{manuel5}
\end{eqnarray}
Varying the action with respect to vierbein $e^i_{\mu}$, the field equations are obtained as 
\begin {equation}\label{manuel6}
e^{-1}\partial_{\mu}(e S^{\:\:\:\mu
\nu}_{i})(1+g_T)-e_i^{\:\lambda}T_{\:\:\:\mu
\lambda}^{\rho}S^{\:\:\:\nu \mu}_{\rho}g_T +S^{\:\:\:\mu
\nu}_{i}\partial_{\mu}(T)g_{TT}-\frac{1}{4}e_{\:i}^{\nu}
(1+g(T))=\frac{1}{2} e_i^{\:\rho}\mathcal{T}_{\rho}^{\:\:\nu},
\end{equation}
where $g_T$ and $g_{TT}$ are the first and second derivatives of $g$ with respect to $T$. Here, $\mathcal{T}_{\rho\nu}$ is the stress tensor and may not be confused with the torsion. Now, we assume the usual
spatially-flat metric of Friedmann-Robertson-Walker (FRW)
universe, with the line element written as
\begin {equation}\label{manuel7}
ds^{2}=dt^{2}-a(t)^{2}\sum^{3}_{i=1}(dx^{i})^{2},
\end{equation}
where $a(t)$ is the scale factor as a function of
the cosmic time $t$. Moreover, we assume the background to
be a perfect fluid. Using the FRW metric
and the perfect fluid matter in the Lagrangian
(\ref{manuel2}) and the field equations (\ref{manuel6}), we obtain
\begin {eqnarray}
T&=&-6H^2\,\,\,,\label{manuel8}\\
3H^2&=& \rho-\frac{1}{2}g-6H^2g_T \,\,\,,\label{manuel9}\\
-3H^2-2\dot{H}&=& p+\frac{1}{2}g+2\left(3H^2+\dot{H}\right)g_T-24\dot{H}H^2g_{TT}\,\,,\label{manuel10}
\end{eqnarray}
where $\rho$ and $p$ are the energy density and pressure  of ordinary matter content of the universe respectively. The Hubble parameter is $H$ and defined as
$H=\dot{a}/a$, where the ``{\it dot}" denotes the derivative with respect to the cosmic time. \par
We now suggest the correspondence between the holographic dark energy scenario and $f(T)$ dark energy model. The holographic dark energy density $\rho_V$ can be written as \cite{setareft,oliver,chinois}
\begin{eqnarray}\label{manuel12}
\rho_V=\frac{3b^2}{R^2_h}\,\,\,,
\end{eqnarray}
where $b$ is a constant, and $R_h$ is the future event horizon defined by 
\begin{eqnarray}\label{manuel13}
R_h=a(t)\int^{\infty}_{t}\frac{dt^{\prime}}{a(t^{\prime})}\,\,\,,
\end{eqnarray}
which, after a simplest transformation, can be put into the following form
\begin{eqnarray}\label{manuel14}
R_h=a\int^{\infty}_{a}\frac{da}{Ha^2}\,\,\,.
\end{eqnarray}
Through the use of the critical energy density $\rho_{cr}=3H^2$, one may define the dimensionless dark energy as
\begin{eqnarray}\label{manuel15}
\Omega_V=\frac{\rho_V}{\rho_{cr}}=\frac{b^2}{H^2R^2_h}\,\,\,.
\end{eqnarray}
Using the definition $\Omega_V$ and $\rho_{cr}$, one gets 
\begin{eqnarray}
\dot{R}_h&=&HR_h-1\,\,\,\,,\nonumber\\
&=& \frac{b}{\sqrt{\Omega_V}}-1\,\,\,.\label{manuel16}
\end{eqnarray}
We assume in this work that the dark energy is the dominant component of the universe and then, it can evolve according to the conservation law
\begin{eqnarray}\label{manuel17}
\dot{\rho}_{V}+3H\left(\rho_V+p_V\right)=0\,\,\,.
\end{eqnarray}
From (\ref{manuel16}) and (\ref{manuel17}), we easily write the derivative of the holographic energy density as
\begin{eqnarray}\label{manuel18}
\dot{\rho}_V=-\frac{-2}{R_h}\left(\frac{b}{\sqrt{\Omega_V}}-1\right)\rho_V\,\,\,,
\end{eqnarray}
from which, using (\ref{manuel17}), one gets
\begin{eqnarray}\label{manuel19}
\omega_V=-\left(\frac{1}{3}+\frac{2\sqrt{\Omega_V}}{3b}\right)\,\,\,.
\end{eqnarray}

We can easily observe that when $\Omega_V\rightarrow 1$ in the future (meaning that holographic dark energy is quite the dominant component of the universe), 
for $b>1$, $\omega_V$ is always greater than $-1$ and
behaves like a quintessence; for $b=1$ the universe will end up with a de Sitter phase, and for $b<1$, the universe falls into a phantom phase and the equation of state crosses $-1$. Therefore, the parameter $b$ plays an important role in specifying the evolutionary nature of the holographic dark energy.\par
The equations (\ref{manuel9}) and (\ref{manuel10}) can be rewritten in order to provide that holographic dark energy corresponds to energy which originates from the deviation part $g(T)$, 
\begin{eqnarray}
3H^2&=&\rho+\rho_V\,\,\,,\quad \quad \rho_V=-\frac{1}{2}g-6H^2g_T\,\,\,,\label{jonas1}\\
-3H^2-2\dot{H}&=& p+p_V\,\,\,,\quad \quad p_V=   \frac{1}{2}g+2\left(3H^2+\dot{H}\right)g_T-24\dot{H}H^2g_{TT}\,\,\,.\label{jonas2}
\end{eqnarray}
Combining (\ref{jonas1}) and (\ref{jonas2}), one obtains the following equation  
\begin{eqnarray}\label{manuel20}
\rho_V+p_V = 2\dot{H}g_T-24\dot{H}H^2g_{TT}\,\,\,.
\end{eqnarray} 
Assuming the equation of state $p_V=\omega_V\rho_V$ for the holographic dark energy, one can rewrite (\ref{manuel20}) as
\begin{eqnarray}\label{manuel21}
-2H^2\Omega_V\left(1-\frac{\sqrt{\Omega_V}}{b}\right)=2\dot{H}g_T-24\dot{H}H^2g_{TT}\,\,\,.
\end{eqnarray} 
Our task now is to determine the model of $f(T)$ coming from holographic dark energy. To do this, let us assume that the Hubble parameter can be written as 
\begin{eqnarray}\label{manuel22}
H(t)= h\left(t_s-t\right)^{-\alpha}\,\,\,,
\end{eqnarray}
where $h$ and $\alpha$ are positive constants, and are assumed in this way for providing the acceleration of the universe. Moreover $t_s$ is the probable future singularity finite time, such that $t<t_s$. Note that (\ref{manuel22}) is specific of the singularities of type I (big rip)  and type III which may appear for $\alpha \geqslant 1$ and  $0<\alpha<1$  respectively \cite{bambaG21,gorbunova7}. For the classification of finite-time future singularities, see \cite{bambaG17}. Using (\ref{manuel22}) and (\ref{manuel8}), one has 
\begin{eqnarray}\label{manuel23}
\dot{H}=\alpha h\left[-\frac{T}{6h^2}\right]^{\frac{\alpha+1}{2\alpha}}\,\,\,,
\end{eqnarray} 
with which we rewrite Eq.(\ref{manuel21}) as 
\begin{eqnarray}\label{manuel24}
2Tg_{TT}+g_T-\frac{1}{\alpha h^3}\Omega_V\left(1-\frac{\sqrt{\Omega_V}}{b}\right)\left[-\frac{T}{6h^2}\right]^{\frac{\alpha-1}{2\alpha}}=0\,\,\,.
\end{eqnarray}
The corresponding scale factor for (\ref{manuel22}) can be written as $a(t)=a_0e^{\frac{h(t_s-t)^{1-\alpha}}{\alpha-1}}$, with which the universe may end up with a future finite time singularity, and one can rewrite (\ref{manuel13}) as
\begin{eqnarray}\label{manuel25}
R_h=a_0e^{\frac{h(t_s-t)^{1-\alpha}}{\alpha-1}}\int_{t}^{t_s}\frac{1}{a_0}e^{-\frac{h(t_s-t^{\prime})^{1-\alpha}}{\alpha-1}}dt^{\prime}\,\,\,.
\end{eqnarray}
In order to let the expression more simplest, we set  $\alpha=1$. Then, Eq.(\ref{manuel25}) becomes
\begin{eqnarray}\label{manuel26}
R_h=\frac{t_s-t}{1+h}\,\,\,,
\end{eqnarray}
from which we get
\begin{eqnarray}\label{manuel27}
\Omega_V=\frac{b^2h^2}{(1+h)^2}\,\,\,,
\end{eqnarray}
and Eq.(\ref{manuel24}) takes the form 
\begin{eqnarray}\label{manuel28}
2Tg_{TT}+g_T+K=0\,\,\,,
\end{eqnarray}
where $K$ is a constant depending on $h$ and $b$ as
\begin{eqnarray}\label{manuel29}
K=-\frac{b^2h^3}{(1+h)^3}\,\,\,.
\end{eqnarray}
The general solution of (\ref{manuel28}) is
\begin{eqnarray}\label{manuel30}
g(T)=-K T+2C_1 \sqrt{-T}+C_2\,\,\,,
\end{eqnarray}
and the corresponding $f(T)$ gravity model according to the holographic dark energy is 
\begin{eqnarray}\label{manuel31}
f(T)=\left(1-K\right)T+ 2C_1 \sqrt{-T}+C_2\,\,\,,
\end{eqnarray}
where $C_1$ and $C_2$ are constants. It is quite evident that at the early time, that we denote $t_0$, the corresponding initial value of the torsion scalar is $T_0$ \footnote{It is easy to obtain this value through Eq. (\ref{manuel8}) in term of the initial Hubble parameter $H_0$ according to the observational data}, 
and we have
\begin{eqnarray}\label{manuel32}
\left(\frac{dT}{dt}\right)_{t=t_0}=-12h^2\left(-\frac{T_0}{6h^2}\right)^{\frac{3}{2}}\,\,\,.
\end{eqnarray}
For a consistency, initial conditions are necessary for determining the respective value of $C_1$ and $C_2$. To do this, we assume the same assumption as in $f(R)$ gravity \cite{chinois}, where, by analogy, the function $f(T)$ has to obeys the following initial conditions
\begin{eqnarray}\label{manuel31}
\left(f\right)_{t=t_0}=T_0\,\,\,, \quad\quad \left(\frac{df}{dt}\right)_{t=t_0}=\left(\frac{dT}{dt}\right)_{t=t_0}\,\,\,\,.
\end{eqnarray} 
Making use of these initial conditions, one obtains
\begin{eqnarray}\label{manuel32}
C_1= K\sqrt{-T_0}\,\,\,,\quad\quad C_2=-KT_0\,\,\,.
\end{eqnarray}
We can then write the explicit expression of $f(T)$ as 
\begin{eqnarray}
f(T)=\left(1-K\right)T+2K\sqrt{T_0T}-KT_0\,\,\,.
\end{eqnarray}
Observe here that when holographic dark energy contribution is almost null (or cancelled), i.e. ($b=0$,\,or $K=0$), $f(T)=T$ (the teleparallel gravity), which would be the model of the matter dominated phase. This result is quite similar to that obtained by S. Nojiri and S. D. Odintsov in the framework of $f(R)$ gravity \cite{matterdominatedphase}, where they investigated the transition of matter dominated phase to the acceleration phase, and found that the matter dominated phase is governed by $f(R)=R$ (Einstein gravity). This may be view as a sort of similarity between the General Relativity and the teleparallel gravity, at least about the matter dominated epoch of the universe.\par
The expression $f(T)=T$, naturally incorporates  the initial conditions and shows the consistency of the model in characterizing dark matter dominated phase. Thus, this is an interesting result in the sense that, the model seems to provide the unification of the dark matter with the dark energy.

\section{Conclusion}
We presented in this paper the reconstruction of $f(T)$ gravity according to holographic dark energy. $f(T)$ theory is a modification of the teleparallel theory for which the gravitational part is the torsion scalar $T$. The teleparallel theory is an analogy of the General Relativity and it is quite natural to investigate some of aspects currently investigated in $f(R)$ gravity, which is a modification of the General Relativity, in this recent theory. Note that holographic principle is an important aspect which can solve some of the problems of cosmological constant and the dark energy. Within any modified theory of gravity, where some cosmological aspects are described, the physical consistency imposes the determination of the action, in other work, the Lagrangian formulation of such a theory. To do this, we put the function $f(T)$ in a strategical form as $T+g(T)$, and try do found the analytical expression of $g(T)$. This form of $f(T)$ is quite reasonable since it means that $f$ is a sum of the teleparallel term plus a correction function $g(T)$. In such case, it appears clearly that in the case of $g(T)=0$, the teleparallel gravity is recovered. We considered the flat FRW universe and the equation of motion is obtained. The holographic dark energy components are introduced and are assumed to dominate other the ordinary matter ones. Differential equation of $g(T)$ is then established and analytically solved, yielding a polynomial function for $g$ and consequently for $f$. The introduced constants $C_1$ and $C_2$ are correctly determined, based on the same assumption applied in $f(R)$  gravity, from initial conditions. Moreover, the input parameter $b$, coming from the holographic dark energy terms, plays an important role, since it is the only parameter that determines the evolutionary aspect of the holographic dark energy. Hence, we see that in future, where $\Omega\rightarrow 1$, if $b>1$,  the dark energy behaves as a quintessence; for $b=1$, the universe ends up with de Sitter phase, while for $b<1$,  $\omega_V$ crossed $-1$ and  the universe  falls into a phantom phase. Another interesting aspect of the obtained model is that it leads to the unification of the dark matter with the dark energy.\par
However, a lot need to be checked within this particular  $f(T)$ theory for providing more explanation for others aspects of dark energy. Note that since $\omega_V$ may cross $-1$, leading to a phantom universe, future finite time singularity (Big rip or type III singularity) can occur. In general, since quantum effects are important near singularity, due to the presence of higher derivative terms of the Hubble parameter, their impact on these singularities may be investigated. One could also allow to the fluid to possess viscosity and then analyse its effect in avoiding singularities. We propose to investigate these aspects in future works.\par \bigskip

{\bf Acknowledgement}: M. H. Daouda thanks CNPq/TWAS for financial support. M. E. Rodrigues  thanks  UFES for the hospitality during the development of this work. M. J. S. Houndjo thanks  CNPq for partial financial support.



\begin{thebibliography}{90}

\bibitem{bamba1} D. N. Spergel et al. [ WMAP Collaboration], Atrophys. J. Suppl. {\bf 148}, 175 (2003); H. V. Peiris et al. {\it ibid.} {\bf 148}, 213 (2003); D. N. Spergel et al. [WMAP Collaboration], {\it ibid}, {\bf 170}, 377 (2007); E. Komatsu et al. [WMAP Collaboration], {\it ibid}, {\bf 180}, 330 (2009). 

\bibitem{bamba2} S. Perlmutter et al. [SNCP Collaboration], Astrophys. J. {\bf 517}, 565 (1999); A. G. Riess et al. [SNST Collaboration], Astron. J. {\bf 116}, 1009 (1998); P. Astier et al. [SNLS Collaboration], Astron. Astrophys. {\bf 447}, 31 (2006); 
A. G. Riess et al., Astrophys. J. {\bf 659}, 98 (2007).


\bibitem{setareh2} M. Tegmark et al. [SDSS Collaboration], Phys. Rev. D {\bf 69}, 103501 (2004); 
K. Abazajian et al. [SDSS Collaboration], Astron. J. {\bf 128}, 502 (2004);
K. Abazajian et al. [SDSS Collaboration], Astron. J {\bf 129}, 1755 (2005).


\bibitem{chinois5} P. J. E. Peebles and B. Ratra, Rev. Mod. Phys. {\bf 75} (2003) 559; S. M. Carroll, Living Rev. Rel. {\bf 4}, 1 (2001);
S. Weinberg, Rev. Mod. Phys. {\bf 61}, 1 (1989).

\bibitem{chinois6}  G. ’t Hooft, [gr-qc/9310026];
L. Susskind, J. Math. Phys {\bf 36}, 6377 (1995).


\bibitem{setareh6} A. G. Cohen, D. B. Kaplan and A. E. Nelson, Phys. Rev. Lett {\bf 82}, 4971 (1999).

\bibitem{setareh7} P. Horava and D. Minic, Phys. Rev. Lett {\bf 85}, 1610 (2000);
S. D. Thomas, Phys. Rev. Lett. {\bf 89}, 081301 (2002).

\bibitem{setareh8} S. D. H. Hsu, Phys. Lett B {\bf 594}, 13 (2004).

\bibitem{setareh9} M. Li, Phys. Lett. B {\bf 603}, 1 (2004).


\bibitem{setareh19} S. Nojiri, S. D. Odintsov, H. Stefancic, Phys. Rev. D {\bf 74}, 086009 (2006); S. Nojiri, S. D. Odintsov, J. Phys. A {\bf 40}, 6725 (2007); G. Cognola, E. Elizalde, S. Nojiri, S.
D. Odintsov and S. Zerbini, Phys. Rev. D {\bf 75}, 086002 (2007); S. Nojiri and , S. D.
Odintsov, J. Phys. Conf. Ser. {\bf 66}, 012005 (2007).

\bibitem{setareh20} S. Capozziello, Int. J. Mod. Phys. D {\bf 11}, 483 (2002); S. Capozziello, S. Carloni and
A. Troisi, arXiv:astro-ph/0303041; S. M. Carroll, V. Duvvuri, M. Trodden and S.
Turner, Phys. Rev. D {\bf 70}, 043528 (2004).

\bibitem{setareh21}  S. Nojiri and S. D. Odintsov, Phys. Rev. D {\bf 68}, 123512 (2003).

\bibitem{setareh22} S. Nojiri, S. D. Odintsov and M. Sasaki, Phys. Rev. D {\bf 71}, 123509 (2005); S. Nojiri,
S. D. Odintsov and M. Sami, Phys. Rev. D {\bf 74}, 046004 (2006); B. M. N. Carter, I. P.
Neupane, Phys. Lett. B{ \bf 638}, 94 (2006); B. M. N. Carter, and I. P. Neupane, JCAP
{\bf 0606}, 004 (2006); J. W. Moﬀat, and V. T. Toth . arXiv: 0710.0364 [astro-ph].

\bibitem{setareh23} S. Nojiri, S. D. Odintsov, 0707.1941v2 [hep-th]; S. Nojiri, S. D. Odintsov, 0710.1738v2
[hep-th]; G. Cognola, E. Elizalde, S. Nojiri, S. D. Odintsov, L. Sebastiani, S. Zerbini,
0712.4017v1 [hep-th].

\bibitem{setarep1} M. R. Setare, 
Phys.Lett B {\bf 642}, 421-425. (2006).


\bibitem{setarep2} M. R. Setare, 
Phys. Lett B {\bf 654}, 1-6 (2007).


\bibitem{setarep3} M. R. Setare, 
Phys.Lett B {\bf 653}, 116-121 (2007).


\bibitem{setarep4} M. R. Setare, 
Int. J. Mod. Phys D {\bf 17}, 2219-2228 (2008).


\bibitem{setarep5} M. R. Setare, 
Eur. Phys. J. C {\bf 50}, 991-998 (2007).


\bibitem{setareh14} Y. Gong, Phys. Rev. D {\bf 70}, 064029 (2004).

\bibitem{setareh15} H. Kim, H. W. Lee, and Y. S. Myung, Phys. Lett. B {\bf 628}, 11 (2005).

\bibitem{setareh16}  D. F. Torres, Phys. Rev. D {\bf 66}, 043522 (2002).

\bibitem{setareh17} M. R. Setare, Phys. Lett. B {\bf 644}, 99 (2007).

\bibitem{chinois} Xing Wu and Zong-Hong Zhu, Phys. Lett. B {\bf 660},293-298 (2008).

\bibitem{oliver} M. J. S. Houndjo and  Oliver F. Piattella, arXiv:1111.4275 [gr-qc].

\bibitem{oliver34}  D. Pavon, W. Zimdahl, Phys. Lett. B {\bf 628}, 206-210 (2005).

\bibitem{oliver35} S. del Campo, J. C. Fabris, R. Herrera, W. Zimdahl, Phys. Rev. D {\bf 83}, 123006 (2011). 

\bibitem{daouda1ref2} C. Moller, Mat. Fys. Skr. Dan. Vid. Selsk. {\bf 1} no. 10 (1961); C. Pellegrini and J. Plebanski, Mat. Fys. Skr. Dan. Vid. Selsk. {\bf 2} no. 4 (1963); C. Moller, K. Dan. Vidensk. Selsk. Mat. Fys. Skr. {\bf 89}, No. 13
(1978); K. Hayashi and T. Nakano, Prog. Theor. Phys. {\bf 38}, 491 (1967); K. Hayashi, Nuovo Cimento
A {\bf 16}, 639 (1973); K. Hayashi, Phys. Lett. B {\bf 69}, 441 (1977); K. Hayashi and T. Shirafuji, Phys. Rev.
D {\bf 19}, 3524-3553 (1979).

\bibitem{setareft16} R. Ferraro and F. Fiorini, Phys. Rev. D {\bf 75}, 084031 (2007).

\bibitem{setareft17} R. Ferraro and F. Fiorini, Phys. Rev. D {\bf 78}, 124019 (2008).


\bibitem{setareft18} G. Bengochca and R. Ferraro, Phys. Rev. D {\bf 79}, 124019 (2009).

\bibitem{daouda1} M. Hamani Daouda, Manuel E. Rodrigues and M. J. S. Houndjo, Eur. Phys. J. C {\bf 71}, 1817 (2011).
 
\bibitem{daouda2} M. Hamani Daouda, Manuel E. Rodrigues and M. J. S. Houndjo, arXiv:1109.0528v2 [physics.gen-ph].

\bibitem{arbitrebamba1} Kazuharu Bamba, Chao-Qiang Geng, Chung-Chi Lee and Ling-Wei Luo, JCAP {\bf 1101}, 021 (2011). 


\bibitem{arbitrebamba2} Kazuharu Bamba, Chao-Qiang Geng and Chung-Chi Lee, 	arXiv:1008.4036v1 [astro-ph.CO].


\bibitem{setareft19} E. V. Linder, Phys. Rev. D {\bf 81}, 127301 (2010).

\bibitem{setareft20} Y-Fu Cai, S-Hung Chen, J. B. Dent, S. Dutta, and E. N. Saridakis, arXiv: 1104.4349v2.

\bibitem{setareft} M. R. Setare and F. Darabi, arXiv: 1110.3962v1 [physics.gen-ph].

\bibitem{setareparticle} M. R. Setare and M. J. S. Houndjo, arXiv:1111.2821 [physics.gen-ph].

\bibitem{bambaG21} K. Bamba, S. Nojiri and S. D. Odintsov, JCAP {\bf 0810}, 045 (2008).

\bibitem{gorbunova7} K. Bamba, S. D. Odintsov, L. Sebastiani and S. Zerbini, Euro. Phys. J. C {\bf 67}, 295-310 (2010). 

\bibitem{bambaG17} S. Nojiri, S. D. Odintsov and S. Tsujikawa, Phys. Rev. D {\bf 71} 063004 (2005).


\bibitem{matterdominatedphase} S. Nojiri and S. D. Odintsov, Phys. Rev. D {\bf 74}, 086005 (2006).

\end{thebibliography}
\end{document}